\documentclass[english,aps,manuscript]{revtex4-1}
\usepackage[T1]{fontenc}
\usepackage[latin9]{inputenc}
\usepackage[active]{srcltx}
\usepackage{float}
\usepackage{amsmath}
\usepackage{amssymb}
\usepackage{graphicx}

\makeatletter

\newcommand{\noun}[1]{\textsc{#1}}

\makeatother

\usepackage{babel}
\begin{document}

\title{Expanding ring-shaped Bose-Einstein condensates as analogs of cosmological
models: Analytical characterization of the inflationary dynamics}

\author{J.M. Gomez Llorente and J. Plata}

\address{Departamento de F\'{\i}sica, Universidad de La Laguna,\\
 La Laguna E38204, Tenerife, Spain.}
\begin{abstract}
We analytically study the expansion of a Bose-Einstein condensate
in a ring-shaped trap with an increasing central radius. The evolution
of the ground state is described using a scaling transform. Additionally,
the dynamics of excited azimuthal modes over the varying \emph{ground}
state is analyzed through a generalization of the Bogoliubov\textendash de
Gennes approach. Our results explain some of the features observed
in recent experiments focused on testing the applicability of the
system as a parallel of cosmological inflationary models. The radial
dynamics, which corresponds to the inflaton field of the cosmological
counterpart, is analytically characterized: the expansion is found
to induce the oscillatory displacement of the condensate as well as
the coupled variation of the radial and vertical widths. Our findings
account also for the observed redshift and emergence of the Hubble
\emph{friction} in the evolution of initially-prepared azimuthal modes.
Our description, which traces the role of the different components
of the setup in the expansion, enhances the controllability, and,
therefore, the potential of the system as a ground for emulating the
inflationary dynamics of cosmological models.
\end{abstract}
\maketitle

\section{Introduction}

Bose-Einstein condensates (BECs) have been proposed as a controllable
scenario for realizing analogs of fundamental effects primarily linked
to other physical contexts. In particular, the possibility of describing
the dynamics of excitations in BECs using equations derived from an
effective curved space-time metric \cite{key-Barcelo,key-Garay1,key-Garay2,key-Gardiner2,key-Gardiner1,key-fedichev}
has allowed the connection with cosmological effects. The difficulties
in observing those effects in their original environment can be avoided
in BECs, where they can be tested under laboratory conditions. Indeed,
the controlled variation of the trap frequencies, interaction strength,
dimensionality, or temperature has made possible the implementation
of different simulation schemes. Significant advances have been achieved
recently. A remarkable example is the use of a BEC to set up a sonic
parallel of a gravitational black hole where the acoustic counterpart
of Hawking radiation \cite{key-Hawking1,key-Unruh,key-Finazzi1,key-Carusotto3}
can be detected \cite{key-Steinhauer1,key-Steinhauer2}. Also noticeable
is the analog of the Sakharov oscillations \cite{key-sakharov} realized
through the interference of acoustic waves generated by appropriately
varying the interaction strength in a \emph{two-dimensional} BEC \cite{key-gurarie}.
Moreover, a closely related mechanism has been used to implement the
dynamical Casimir effect \cite{key-nationRMP}: quasi-particles have
been created from the \emph{vacuum} by suitably modifying the transversal
trap frequencies in an elongated condensate \cite{key-jaskula}. Here,
we focus on recent work which can open another line in this field
of \emph{cold-atom cosmology}, namely, the proposal for setting up
a parallel of inflationary universe models\emph{ }presented in Ref.
\cite{key-SpielmanRing}. Specifically, it has been proposed that
the evolution of an atomic BEC in a ring trap with a growing central
radius can simulate some of the characteristics of the primordial
inflation in cosmology. The practical realization has shown features
that substantiate the analogy \cite{key-SpielmanRing}. The increase
of the trap radius has been observed to alter the radial dynamics
in a way reminiscent of the forms predicted for the inflaton field,
i.e., for the field conjectured to drive the cosmological expansion
\cite{key-amin,key-starobinsky1,key-starobinsky2,key-felder,key-figueroa}.
Additionally, results of numerical simulations, to be experimentally
verified, have uncovered a scenario that seems to emulate the \emph{reheating}
process, i.e., the thermalization predicted in the cosmological context.
After the expansion, the energy stored in the radial excitations is
found to induce the generation of solitons, paralleling the \emph{preheating}
phase, where the inflaton field leads to the production of particles.
The solitons, in turn, decay into vortices, which subsequently bring
about the emergence of stochastic currents. The eventual thermalization,
although not proved, seems to be reasonably conjectured. The parallelism
 with the cosmological setting can also be traced in the detected
dynamics of initially-prepared azimuthal modes: the redshift of the
mode frequency and a counterpart of the ``Hubble friction'' were
uncovered. Our work is intended to clarify those analogies by giving
an analytical description of the dynamics. Our theoretical framework,
based on a scaling transform for the ground state and on a generalization
of the Bogoliubov\textendash de Gennes (BdG) approach for the excitations,
allows dealing with a general regime of time variation in the evolution
of the condensate and incorporates corrections to the phononic term
in the dynamical dispersion relation of the modes. Consequently, we
can account for effects like the displacement of the condensate with
respect to the trap or the coupling of the radial and vertical dynamics
in the expansion. We can also evaluate the role of the background
nonadiabaticity in the mode evolution, or trace the origin of features
observed in the redshifting and the ``Hubble friction''. Our results
can pave the way for designing elements to control the involved processes,
and, consequently, for advancing in the characterization of  the referred
effects in laboratory conditions.

The outline of the paper is as follows. In Sec. II, we present our
model system. A scaling method is applied in Sec. III to analyze the
 evolution of the condensate  during and after the expansion of the
trap. In Sec. IV, the dynamics of excited azimuthal modes is studied.
In order to deal with the time-dependent background, we set up a generalization
of the BdG approach. Moreover, the emergent redshift and Hubble \emph{friction}
are characterized in Sec. V using an effective Hamiltonian description
of the modes. In Sec. VI, we evaluate the appearance of effects specific
to the anharmonicity of the trap used in the experimental realization
\cite{key-SpielmanRing}.  Finally, some general conclusions are summarized
in Sec. VII.

\section{The model system}

We consider an atomic BEC in a ring-shaped confining potential $V_{ex}(\mathbf{r},t)$
whose central radius is made to vary. Specifically, in cylindrical
coordinates ($r$, $\theta$, $z$), the trapping potential is given
by 

\begin{equation}
V_{ex}(r,z,t)=\frac{1}{2}M\omega_{z}^{2}z^{2}+\frac{1}{2}M\omega_{r}^{2}[r-R(t)]^{2}+\Lambda[r-R(t)]^{4},\label{eq:ExternalPotential}
\end{equation}
 where $M$ is the mass of a condensate atom, $\omega_{z}$ and $\omega_{r}$
denote respectively the vertical and radial frequencies, $\Lambda$
accounts for the anharmonicity of the trap in the radial direction,
and $R(t)$ stands for the time-dependent central radius of the ring.
In the experimental realization \cite{key-SpielmanRing}, $R(t)$
was made to steadily grow in order to parallel the expansion in an
inflationary universe model. As corresponds to the characteristics
of the practical implementation, the potential does not depend on
the azimuthal coordinate $\theta$. No limitations are assumed on
the relative magnitude of the radial and vertical frequencies. The
experimental arrangement, which corresponds to $\omega_{z}$ being
significantly larger than $\omega_{r}$, will be analyzed as a particular
case in our general framework. The effect of the anharmonic term $\Lambda[r-R(t)]^{4}$
will be analyzed in Sec. VI; till then, a purely harmonic confinement
will be considered.

The mean-field description of the system wave-function $\Psi(\mathbf{r},t)$
is given by the time-dependent three-dimensional Gross-Pitaevskii
(GP) equation \cite{key-Stringari}

\begin{equation}
i\hbar\frac{\partial\Psi(\mathbf{r},t)}{\partial t}=\left[-\frac{\hbar^{2}}{2M}\nabla^{2}+V_{ex}(\mathbf{r},t)+g\left|\Psi(\mathbf{r},t)\right|^{2}\right]\Psi(\mathbf{r},t),\label{eq:TimeDepGP}
\end{equation}
 where $g$ is the strength that characterizes the atom-atom interaction. 

Convenient for the implementation of our analytical approach is the
use of the hydrodynamic formalism. Applying it in the Thomas-Fermi
(TF) regime, we derive a system of equations for the density $\rho(\mathbf{r},t)$
and velocity field $\mathbf{v}(\mathbf{r},t)$ from the time-dependent
GP equation. Namely, rewriting the wave-function in terms of its modulus
and phase, i.e., 

\begin{equation}
\Psi(\mathbf{r},t)=\sqrt{\rho(\mathbf{r},t)}e^{iS(\mathbf{r},t)},\label{eq:hydroAnsatz}
\end{equation}
 and with
\begin{equation}
\mathbf{v}(\mathbf{r},t)=\frac{\hbar}{M}\boldsymbol{\mathbf{\nabla}}S(\mathbf{r},t),\label{eq:VelocityPhase}
\end{equation}
 we derive the  continuity and the Euler-like (hydrodynamic) equations,
which respectively read 

\begin{equation}
\frac{\partial\rho}{\partial t}+\boldsymbol{\nabla}(\rho\mathbf{v})=0\label{eq:Continuity}
\end{equation}

\begin{equation}
M\frac{\partial\mathbf{v}}{\partial t}=\boldsymbol{\nabla}\left(-\frac{1}{2}M\mathbf{v}^{2}-V_{ex}-g\rho\right).\label{eq:Euler}
\end{equation}
 The applicability of the TF approximation to describe the ground-state
dynamics will be assumed throughout the analytical part of the study.
We have numerically checked that, given the experimental conditions,
this is a sound assumption.

Following the practical procedure, we will consider two different
preparations. First, the system will be assumed to be in the ground
state corresponding to the external potential previous to the expansion,
i.e., to $V_{ex}(\mathbf{r},t)$ with the central radius taking its
initial value $R(t=0)\equiv R_{0}$. The evolution resulting from
the variation of $R(t)$ will be analytically characterized. Second,
it will be considered that, initially, an azimuthal mode has been
excited from the previously described ground state. We will analyze
how the corresponding perturbation in the density and phase of the
condensate evolves during the expansion of the trap.

\section{Analytical description of the evolution of the ground state}

\subsection{Application of a scaling approach}

From the experimental results presented in Ref. \cite{key-SpielmanRing},
it is apparent that an adiabatic approximation is not applicable:
the evolution of the condensate during the trap expansion differs
from the sequence of static ground-state configurations corresponding
to \emph{frozen }values of\emph{ }the time-varying central radius.
To explain the experimental findings, we must set up a theoretical
framework valid in a general regime of time variation. We will start
with a simplified scenario where a purely harmonic radial confinement
is considered. It will be shown that the mechanisms responsible for
the radial dynamics during the expansion can be already identified
in this description. Subsequently, effects specific to the radial
anharmonicity will be evaluated. 

Our approach consists in a variant of the scaling methods applied
in previous studies of condensates in harmonic traps with time dependent
frequencies \cite{key-Castin}\cite{key-Kagan}. We will adapt those
techniques to the present case, where the frequencies take fixed values
and it is the trap central-radius that is varied. Significant differences
with the behavior observed in former setups will be shown to derive
from the ring geometry, specifically, from the radial coordinate being
the direction of the trap expansion. As in previous presentations
of the scaling methods, we start by relating the condensate density
at time $t$, $\rho(r,z,\theta,t)$, with the initial density, $\rho(r_{0},z_{0},\theta_{0},0)$,
through a transformation of the variables: the time evolution is incorporated
in the scaling of the initial coordinates $(r_{0},z_{0},\theta_{0})$
to give the final ones $(r,z,\theta)$. The assumed uniformity in
the azimuthal coordinate implies that the (initially-prepared) ground
state does not depend on $\theta$. Moreover, the uniformity in $\theta$
is maintained in the state evolution. The equations for the transformation
of $r$ and $z$ read 

\begin{eqnarray}
r-R(t) & = & \sigma_{r}(t)(r_{0}-R_{0})+\lambda_{r}(t),\label{Rtransformation}\\
z & = & \sigma_{z}(t)z_{0},\label{eq:Ztransformation}
\end{eqnarray}
 where we have introduced three scaling functions: the \emph{width}
factors, $\sigma_{r}(t)$, $\sigma_{z}(t)$, which account for changes
in the shape of the condensate, and the \emph{displacement} factor
$\lambda_{r}(t)$, which describes a radial translation  with respect
to the trap. (The introduction of an additional \emph{displacement}
$\lambda_{z}(t)$ in the $z$-direction will be shown to be irrelevant
given the characteristics of the considered realization). Note that
$r-R(t)$ corresponds to the co-moving radial coordinate used in the
analysis of the experimental results presented in \cite{key-SpielmanRing}.

Now, from the conservation of the number of particles, ($N$ atoms
are considered), we derive the equation 
\begin{eqnarray}
\rho(r,z,t) & = & \frac{r_{0}}{r}\frac{1}{\sigma_{r}\sigma_{z}}\rho_{0}\left(r_{0},z_{0},0\right),\label{eq:numberNconserv}
\end{eqnarray}
 which, using the scaling transformation, {[}Eqs. (\ref{Rtransformation})
and (\ref{eq:Ztransformation}){]}, is rewritten as

\begin{equation}
\rho(r,z,t)=\frac{1}{r}\left[\frac{r-R(t)-\lambda_{r}}{\sigma_{r}}+R_{0}\right]\frac{1}{\sigma_{r}\sigma_{z}}\rho_{0}\left(\frac{r-R(t)-\lambda_{r}}{\sigma_{r}}+R_{0},\frac{z}{\sigma_{z}},0\right).\label{eq:timedepDensity}
\end{equation}
Note that the presence of the factor $\frac{r_{0}}{r}=\frac{1}{r}(\frac{r-R(t)-\lambda_{r}}{\sigma_{r}}+R_{0})$
is specific to the radial coordinate. We will see that it is because
of this factor that a nontrivial variation of the shape of the condensate,
embodied by the functions $\sigma_{r}(t)$ and $\sigma_{z}(t)$, emerges. 

From the above equation, and, taking as initial density that given
by the Thomas-Fermi approximation, we obtain 

\begin{equation}
\rho(r,z,t)=\frac{1}{r}\left[\frac{r-R(t)-\lambda_{r}}{\sigma_{r}}+R_{0}\right]\frac{\mu-\frac{1}{2}M\omega_{z}^{2}(z/\sigma_{z})^{2}-\frac{1}{2}M\omega_{r}^{2}[(r-R(t)-\lambda_{r})/\sigma_{r}]^{2}}{g\sigma_{r}\sigma_{z}},\label{eq:FinalDensity}
\end{equation}
 where $\mu$ is the chemical potential corresponding to the system
before the expansion. 

Inserting the ansatz given by Eq. (\ref{eq:timedepDensity}) into
the continuity equation, Eq. (\ref{eq:Continuity}), we find for the
velocity field

\begin{equation}
\mathbf{v}=\left[\frac{\dot{\sigma_{r}}}{\sigma_{r}}(r-R-\lambda_{r})+\dot{\lambda_{r}}+\dot{R}\right]\boldsymbol{u}_{r}+\frac{\dot{\sigma_{z}}}{\sigma_{z}}z\boldsymbol{u}_{z},\label{eq:VelocityField}
\end{equation}
 where $\boldsymbol{u}_{r}$ and $\boldsymbol{u}_{z}$ are, respectively,
unitary vectors in the radial and vertical directions. Consequently,
from Eq. (\ref{eq:VelocityPhase}), the condensate phase is shown
to be given by

\begin{equation}
S(r,z,t)=\frac{M}{\hbar}\left[\frac{1}{2}\frac{\dot{\sigma_{r}}}{\sigma_{r}}r^{2}+\left(-\frac{\dot{\sigma_{r}}}{\sigma_{r}}(R+\lambda_{r})+\dot{\lambda_{r}}+\dot{R}\right)r+\frac{1}{2}\frac{\dot{\sigma_{z}}}{\sigma_{z}}z^{2}\right].\label{eq:FinalPhase}
\end{equation}

The equations for the scaling functions $\sigma_{r}(t)$, $\sigma_{z}(t)$,
and $\lambda_{r}(t)$ are obtained by making the density given by
Eq. (\ref{eq:FinalDensity}) to fulfill the Euler-like equation, Eq.
(\ref{eq:Euler}). A closed set of equations is derived if we make
the approximation 
\begin{equation}
r\simeq r_{c}(t)=\lambda_{r}(t)+R(t)\label{eq:rc-1}
\end{equation}
 in Eq. (\ref{eq:FinalDensity}), with $r_{c}(t)$ being the radial
coordinate of the maximum of the density $\rho(r,z,t)$. That approximation
is indeed valid when the ring width is much smaller than the central
radius. Because of the expansion, if this restriction is satisfied
at the initial time, it is also fulfilled at any subsequent time.
Accordingly, we find for the scaling functions

\begin{eqnarray}
\ddot{\sigma_{r}} & = & -\omega_{r}^{2}\left(\sigma_{r}-\frac{1}{\sigma_{r}^{2}\sigma_{z}}\frac{R_{0}}{r_{c}(t)}\right)\label{eq:sigmaREq}\\
\ddot{\sigma_{z}} & = & -\omega_{z}^{2}\left(\sigma_{z}-\frac{1}{\sigma_{r}\sigma_{z}^{2}}\frac{R_{0}}{r_{c}(t)}\right)\label{eq:sigmaZEq}\\
\ddot{\lambda_{r}} & = & -\omega_{r}^{2}\lambda_{r}-\ddot{R}.\label{eq:lambdaR}
\end{eqnarray}
 There is a first set of initial conditions $\sigma_{r}(0)=1$, $\sigma_{z}(0)=1$,
$\lambda_{r}(0)=0$, which derive from the form of the density of
the prepared state. Moreover, since, at $t=0$, the condensate is
at rest, we must add $\dot{\sigma_{r}}(0)=0$, $\dot{\sigma_{z}}(0)=0$,
and $\dot{\lambda_{r}}(0)=-\dot{R}$. With these equations, a null
velocity field at $t=0$ is consistently obtained from Eq. (\ref{eq:VelocityField}).
Variations in the experimental realization can be incorporated into
our framework by appropriately modifying the initial conditions. 

Throughout our study we have checked the validity of the scaling-approach
predictions by comparing them with the results of a simulation based
on numerically solving the GP equation. In particular, we have confirmed
the applicability of the Thomas-Fermi approximation. The system characteristics
have allowed applying the techniques presented in Ref. \cite{key-salasnich}
for simplifying the numerical resolution of the GP equation. Indeed,
as the vertical frequency is significantly larger than the radial
one and there is no confinement in the azimuthal direction, the vertical
dynamics is incorporated through effective parameters in the equation
for the radial and azimuthal coordinates. Specifically, adapting the
variational approach of Ref. \cite{key-salasnich} to the ring geometry,
we have derived a non-polynomial nonlinear Schrödinger equation applicable
to the present context. Namely, we have used the ansatz 

\begin{equation}
\Psi(\mathbf{r},t)=\phi(r,t)f\left(z,t;\varsigma(r,t)\right)\label{eq:NonPolyAnsatz}
\end{equation}
 where $f\left(z,t;\varsigma(r,t)\right)$ is a Gaussian of width
$\varsigma(r,t)$. The corresponding Euler-Lagrange equations are
straightforwardly obtained; making the functional change $\phi(r,t)=r^{-1/2}\varPhi(r,t)$,
they read 

\begin{eqnarray}
i\hbar\frac{\partial\varPhi(r,t)}{\partial t} & = & \biggl[-\frac{\hbar^{2}}{2M}\frac{\partial^{2}}{\partial r^{2}}+V_{ex}(\mathbf{r},t)-\frac{\hbar^{2}}{4M}\frac{1}{r^{2}}+\nonumber \\
 &  & \frac{g}{(2\pi)^{1/2}}\frac{\left|\varPhi(\mathbf{r},t)\right|^{2}}{r\varsigma(r,t)}+\frac{\hbar^{2}}{2M}\frac{1}{\varsigma^{2}(r,t)}+\frac{1}{2}M\omega_{z}^{2}\varsigma^{2}(r,t)\biggr]\varPhi(r,t)\label{eq:NonPolyn}
\end{eqnarray}

\begin{equation}
\varsigma^{4}(r,t)-2(2\pi)^{1/2}\left(\frac{\hbar}{M\omega_{z}}\right)^{2}a_{s}\frac{\left|\varPhi(\mathbf{r},t)\right|^{2}}{r}\varsigma(r,t)-\left(\frac{\hbar}{M\omega_{z}}\right)^{2}=0,\label{eq:NonPolyWidth}
\end{equation}
 where $a_{s}=\frac{M}{4\pi\hbar^{2}}g$ is the scattering length.
Additionally, the boundary condition $\varPhi(0,t)=0$ and the normalization
requirement $2\pi\int_{0}^{\infty}\left|\varPhi(\mathbf{r},t)\right|^{2}dr=N$
must be imposed. Note that the form of Eq. (\ref{eq:NonPolyn}) allows
the application of standard split-operator techniques of integration.
In our calculations, the initial state of the system has been obtained
through imaginary-time propagation.

\subsection{Characterization of the inflationary dynamics}

The equations obtained for the scaling functions provide the following
clues to the system behavior:

i) The experimentally observed departure of the ground-state evolution
from the adiabatic regime can be precisely characterized in our approach.
Eq. (\ref{eq:lambdaR}) uncovers that the radial displacement of the
condensate with respect to the trap corresponds to a harmonic oscillator
driven by the acceleration of the trap central-radius. The analysis
is particularly simple for a linear ramp: since the driving term in
Eq. (\ref{eq:lambdaR}) disappears ($\ddot{R}=0$), it follows that
the condensate, for the considered initial conditions, oscillates
around the trap inner-center position with the radial frequency $\omega_{r}$.
In that case, it is the initial velocity $\dot{\lambda_{r}}(0)=-\dot{R}$
that generates the oscillation; the energy associated with the radial
displacement is then a constant determined by the ramp velocity. In
the experimental realization, a nonlinear ramp, designed to smoothly
initiate and end the time variation of $R$, was used. The numerical
study of the corresponding displacement, obtained by including the
specific functional form of $R(t)$ in Eq. (\ref{eq:lambdaR}), shows
that the oscillatory behavior is still present. This is apparent in
Fig. 1, where we depict the radius profile $R(t)$ along with the
radial position of the maximum of $\rho(r,z,t)$, $r_{c}(t)$. In
agreement with the experimental findings, our results show that the
condensate is initially delayed and subsequently advanced with respect
to the trap center. It is also observed that the times of expansion
in the experiment are not sufficiently large for observing a complete
oscillation during the ramp. (Note that the complete cycles in Fig.
1 correspond to the post-inflationary stage).

\begin{figure}[H]
\centerline{\includegraphics{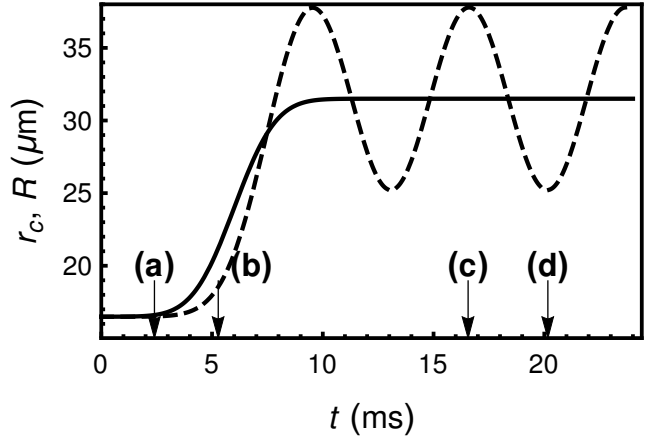}}\caption{The trap radius $R(t)$ (continuous line) and the radial position
of the maximum of the condensate density $r_{c}(t)$ (dashed line)
as functions of time. The form of the radius ramp is the same as that
used in \cite{key-SpielmanRing}. The arrows correspond to times shown
in Fig. 2. The system parameters are $\omega_{r}=2\pi\times200\,\textrm{Hz}$,
$\omega_{z}=2\pi\times650\,\textrm{Hz}$, $N=2\times10^{5}$. ($N$
is the number of atoms).}
\end{figure}

When setting up the scaling transform, we considered that the displacement
of the condensate with respect to the trap takes place only in the
radial direction. This is not a limitation of our model but a simplification
allowed by the experimental arrangement. Actually, the addition of
a function $\lambda_{z}(t)$ to the $z$-transformation {[}Eq. (\ref{eq:Ztransformation}){]}
is irrelevant: one trivially finds $\lambda_{z}(t)=0$ for the initial
conditions $\lambda_{z}(0)=0$ and $\dot{\lambda_{z}}(0)=0$, which
correspond to the practical realization, where no translation of the
trap in the vertical direction was arranged. 

ii) Features specific to the trap geometry, and, in particular, to
the fact that the driving takes place in the radial coordinate can
be identified in the equations for $\sigma_{r}(t)$ and $\sigma_{z}(t)$.
Actually, for a harmonic trap in Cartesian coordinates, since there
is no factor like $\frac{r_{0}}{r}$ in the counterpart of Eq. (\ref{eq:numberNconserv}),
the scaling functions $\sigma_{i}(t)$ ($i=x,\,y,\,z$) obey equations
similar to those obtained for the ring trap with the factor $\frac{R_{0}}{r_{c}(t)}$
replaced by $1$. Then, for a realization where, starting from the
condensate at rest, the trap center is displaced, one trivially finds
$\sigma_{i}(t)=1$ ($i=x,\,y,\,z$). Hence, in that setup, there is
no change in the shape of the condensate: the dynamics is solely given
by the translation functions $\lambda_{i}(t)$.

In Fig. 2, we present the effective radial density at four different
times for parameters similar to those used in the experiments. Some
features of the deformation can be identified in it. Note that the
significant differences in the heights of the density at the different
times result basically from the normalization condition in the radial
coordinate. It is worth stressing that there is almost complete agreement
between the analytical results obtained with the scaling approach
and those found in the numerical simulation. The radial deformation
incorporated by $\sigma_{r}(t)$ is hardly visible in Fig. 2. However,
a spectral analysis uncovers the existence of nontrivial changes in
the widths of the condensate. As it can be useful for a proposal of
experimental detection of those changes, we rewrite the condensate
density given by Eq. (\ref{eq:FinalDensity}) in terms of effective
time-dependent frequencies. Specifically, the density is expressed
as 

\begin{equation}
g\rho(r,z,t)=\frac{R_{0}}{r_{c}(t)\sigma_{r}\sigma_{z}}\mu-\frac{1}{2}M\omega_{z,eff}^{2}(t)z^{2}-\frac{1}{2}M\omega_{r,eff}^{2}(t)(r-r_{c})^{2},
\end{equation}
 where 
\begin{equation}
\omega_{r,eff}(t)=\omega_{r}\frac{1}{\sigma_{r}}\sqrt{\frac{R_{0}}{r_{c}\sigma_{r}\sigma_{z}}}\qquad\omega_{z,eff}(t)=\omega_{z}\frac{1}{\sigma_{z}}\sqrt{\frac{R_{0}}{r_{c}\sigma_{r}\sigma_{z}}}.
\end{equation}
 Now, we can think of using the analysis of the frequency shifts defined
by $\Delta\omega_{r}(t)=\omega_{r,eff}(t)-\omega_{r}$ and $\Delta\omega_{z}(t)=\omega_{z,eff}(t)-\omega_{z}$
as an operative form of evaluating the condensate deformation. The
nontrivial evolution of $\Delta\omega_{r}(t)$ and $\Delta\omega_{z}(t)$,
rooted in the coupling of $\sigma_{r}(t)$ and $\sigma_{z}(t)$, is
illustrated in Figs. 3 and 4. The much smaller magnitude of the deformation
in the $z$-coordinate is apparent. The values of $\Delta\omega_{r}(t)/\omega_{r}$
imply relative changes of at most $\sim0.4\times10^{-2}$.

\begin{figure}[H]
\centerline{\includegraphics{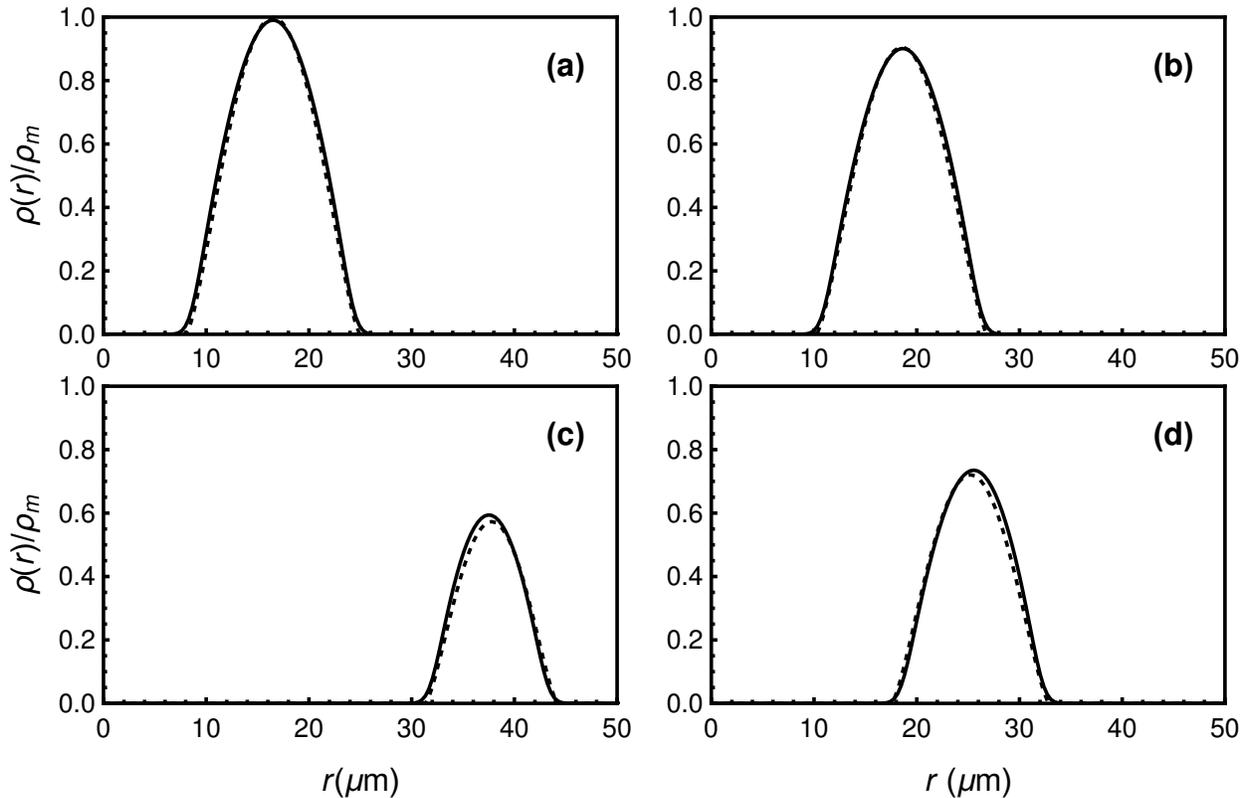}}\caption{The reduced radial density $\rho(r)/\rho_{m}$ at four different times.
($\rho_{m}$ denotes the maximum of $\rho(r)$ before the expansion).
The labels $a$, $b$, $c$, and $d$ correspond to the times identified
in Fig. 1. The dashed lines represent the results obtained analytically
with the scaling approach. The continuous lines correspond to the
results obtained by numerically solving the non-polynomial nonlinear
Schrödinger  equation. The system parameters are the same as those
used in Fig. 1.}
\end{figure}

\begin{figure}[H]
\centerline{\includegraphics{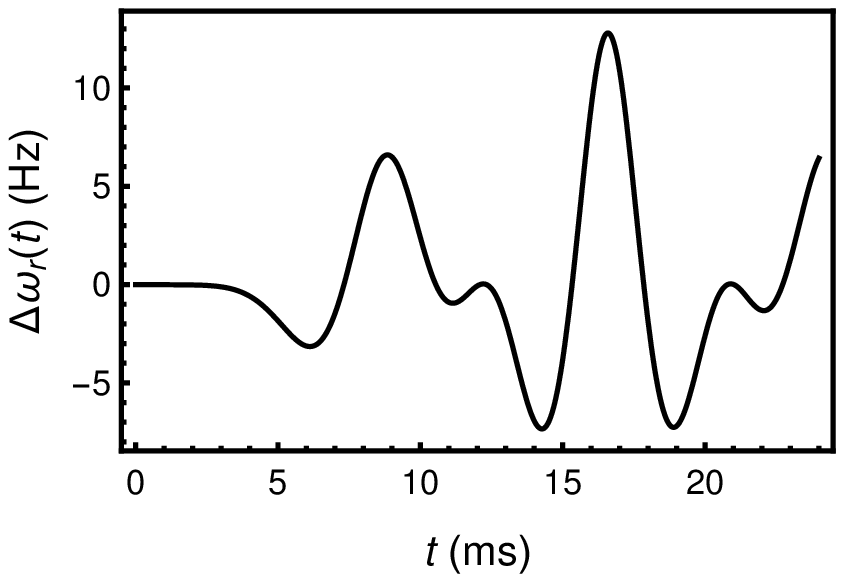}}\caption{The effective dynamical shift in the radial frequency $\Delta\omega_{r}(t)=\omega_{r}\left(\frac{1}{\sigma_{r}}\sqrt{\frac{R_{0}}{r_{c}\sigma_{r}\sigma_{z}}}-1\right)$
as obtained from our analytical scaling approach. The system parameters
are the same as those used in Fig. 1.}
\end{figure}

\begin{figure}[H]
\centerline{\includegraphics{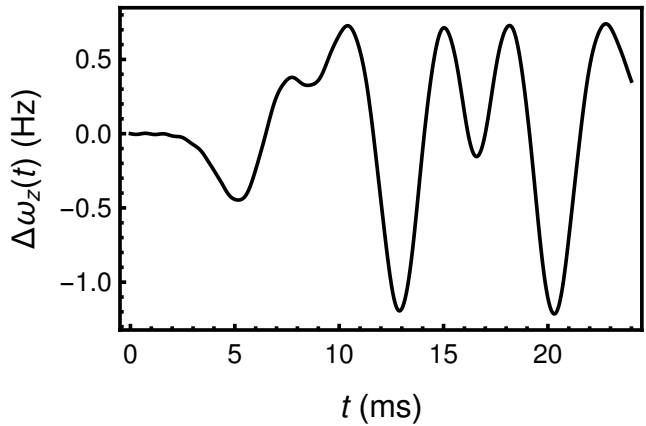}}\caption{The effective dynamical shift in the radial frequency $\Delta\omega_{z}(t)=\omega_{z}\left(\frac{1}{\sigma_{z}}\sqrt{\frac{R_{0}}{r_{c}\sigma_{r}\sigma_{z}}}-1\right)$
as obtained from our analytical scaling approach. The system parameters
are the same as those used in Fig. 1.}
\end{figure}

iii) Important for tracing parallelisms with the cosmological \emph{reheating}
is the characterization of the system behavior once the expansion
stops. Eq. (\ref{eq:lambdaR}) predicts that, after the inflation,
the condensate oscillates with the radial frequency. This is observed
in Fig. 1. The persistence of this (regular) oscillation, which can
be regarded as an effective radial dipole mode, is specific to the
considered case of a purely harmonic confinement. As shown in Sec.
VI, a more complex behavior emerges when the anharmonic term is incorporated
into the description. Indeed, the trap anharmonicity present in the
practical setup can be identified as one of the factors responsible
for the complex dynamics detected in the numerical simulation presented
in Ref. \cite{key-SpielmanRing}. 

iv) The nonlinear character of the equations for the width factors
persists after the expansion. A picture of the condensate deformation
can be obtained from Figs. 3 and 4. As the factor $\frac{R_{0}}{r_{c}(t)}$
has decreased during the ramp, a more regular behavior emerges in
the post-inflationary stage. Actually, a spectral analysis reveals
the quasi-periodic character of the outputs in $\Delta\omega_{r}(t)$
and $\Delta\omega_{z}(t)$. Some clues to the dominant frequencies
 can be extracted from the linearization of the system of Eqs. (\ref{eq:sigmaREq})
and (\ref{eq:sigmaZEq}). Assuming that the expansion ends at $t_{f}$,
{[}$R(t)=R_{f}$ for $t\geq t_{f}${]}, we take $\sigma_{r}\simeq\sigma_{r,f}+\delta\sigma_{r}$
and $\sigma_{z}\simeq\sigma_{z,f}+\delta\sigma_{z}$, where $\sigma_{r,f}=\sigma_{z,f}=(R_{0}/R_{f})^{1/4}$
are the \emph{equilibrium} values of $\sigma_{r}$ and $\sigma_{z}$
at the end of the expansion. Then, working to first order in $\delta\sigma_{r}$
and $\delta\sigma_{z}$, one finds the equations 

\begin{eqnarray}
\ddot{\delta\sigma_{r}} & = & -\omega_{r}^{2}\left(3\delta\sigma_{r}+\delta\sigma_{z}\right)\label{eq:DeltaSigmaR}\\
\ddot{\delta\sigma_{z}} & = & -\omega_{z}^{2}\left(\delta\sigma_{r}+3\delta\sigma_{z}\right),\label{eq:DeltaSigmaZ}
\end{eqnarray}
 which (still) display the coupling of the radial and vertical widths
in the associated normal modes. It is possible to go further analytically
when the characteristic times are widely different. Let us consider
the case corresponding to $\omega_{z}$ being significantly larger
than $\omega_{r}$, which is close to the experimental realization.
As shown in the Appendix, in that situation, an effective decoupling
of $\delta\sigma_{r}$ and $\delta\sigma_{z}$ can be shown to occur.
The frequencies of the oscillations in the radial and vertical widths
are then respectively given by $\tilde{\omega}_{r}=\sqrt{8/3}\omega_{r}$
and $\tilde{\omega}_{z}=\sqrt{3}\omega_{z}$. The spectral analysis
of the detected  outputs can provide the data required for the experimental
verification of these results.

Beyond the considered expanding scenario, our approach is applicable
to a general variation of $R$. In particular, it is possible to tackle
the effects of a steady reduction or of sudden changes in the radius.
In the application to a decreasing-radius arrangement, the approximation
$\frac{r_{0}}{r}\simeq\frac{R_{0}}{r_{c}(t)}$ must be reevaluated.
Additionally, since the (nonlinear) coupling of the radial and vertical
widths is determined by the quotient $\frac{R_{0}}{r_{c}(t)}$, it
can be predicted to lead to stronger effects in a contracting scenario. 

The study can be straightforwardly generalized to include changes
in the trap frequencies or an initial displacement in the $z$-direction.
The implications of those variants of the basic setup for the selective
excitation of modes are evident. Actually, useful clues to the design
of specific characteristics in the (atomic) \emph{inflaton} field
can be extracted from the obtained analytical results.

\section{The dynamics of the azimuthal modes}

Apart from tracing the evolution of the fundamental state, the experiments
of Ref. \cite{key-SpielmanRing} were focused on the effect of the
ring expansion on azimuthal modes initially prepared in the condensate.
In the practical procedure for exciting the modes, a perturbation
$\delta V(\theta,t)$ was applied to the external potential. Subsequently,
the unperturbed potential, (uniform in $\theta$), was reestablished
and the expansion of the ring was implemented. In our description
of those processes, we write the condensate wave-function $\Psi(\mathbf{r},t)$
as the sum of that of the evolved ground state $\Psi_{0}(\mathbf{r},t)$,
obtained in the previous section, and the perturbation $\delta\Psi(\mathbf{r},t)$
excited via the modification of the external potential. In order to
directly compare with the (experimental and theoretical) results presented
in Ref. \cite{key-SpielmanRing}, it is convenient to express the
perturbation as $\delta\Psi(\mathbf{r},t)=\Psi_{0}(\mathbf{r},t)\delta\varphi(\mathbf{r},t)$.
Therefore, we write 

\begin{equation}
\Psi(\mathbf{r},t)=\Psi_{0}(\mathbf{r},t)\left[1+\delta\varphi(\mathbf{r},t)\right].\label{eq:PerturbAnsatz}
\end{equation}
 Using this ansatz, the connection with the perturbed density and
phase, which are split as

\begin{equation}
\rho(\mathbf{r},t)=\rho_{0}(\mathbf{r},t)+\delta\rho(\mathbf{r},t)\label{eq:SplittRho}
\end{equation}

\begin{equation}
S(\mathbf{r},t)=S_{0}(\mathbf{r},t)+\delta S(\mathbf{r},t),\label{eq:SplittPhase}
\end{equation}
 is simply given by the equations

\begin{equation}
\delta S(\mathbf{r},t)=\textrm{Im}\left\{ \delta\varphi(\mathbf{r},t)\right\} \label{eq:DeltaPhasePsi}
\end{equation}

\begin{equation}
\delta\rho(\mathbf{r},t)=2\rho_{0}(\mathbf{r},t)\textrm{Re}\left\{ \delta\varphi(\mathbf{r},t)\right\} .\label{eq:DeltaRhoPsi}
\end{equation}
 To characterize the evolution of $\delta\varphi(\mathbf{r},t)$,
we will set up a generalized version of the BdG approach which will
allow us to describe the dynamics of the modes over the time-varying
background.

\subsection{Generalized Bogoliubov\textendash de Gennes  approach}

As in the standard procedure \cite{key-Finazzi1,key-Carusotto3},
we start by defining the column vector 

\begin{equation}
\boldsymbol{\delta\Psi}\equiv\begin{pmatrix}\delta\Psi\\
\delta\Psi^{*}
\end{pmatrix}.\label{eq:ClassicalQuanConV-1}
\end{equation}
Its evolution, obtained through the linearization of the time-dependent
GP equation {[}Eq. (\ref{eq:TimeDepGP}){]}, is  given by 

\begin{equation}
i\hbar\frac{\partial\boldsymbol{\delta\Psi}}{\partial t}=\boldsymbol{\mathcal{L}_{BdG}}\boldsymbol{\delta\Psi}+\boldsymbol{W}.\label{eq:BdGennes}
\end{equation}
 In this equation, two contributions to the dynamics can be differentiated.
First, the homogeneous part, characterized by the BdG operator

\begin{equation}
\boldsymbol{\mathcal{L}_{BdG}}\equiv\begin{pmatrix}-\frac{\hbar^{2}}{2M}\nabla^{2}+V_{ex}+2g\left|\Psi_{0}\right|^{2} & g\Psi_{0}^{2}\\
-g\Psi_{0}^{*2} & -\left(-\frac{\hbar^{2}}{2M}\nabla^{2}+V_{ex}+2g\left|\Psi_{0}\right|^{2}\right)
\end{pmatrix},\label{eq:AdBGOper-2}
\end{equation}
 accounts for the role of the time-dependent substrate in the evolution
of the perturbation. $\boldsymbol{\mathcal{L}_{BdG}}$ presents explicit
time dependence through $\Psi_{0}$ and $V_{ex}$. 

Second, the source matrix 

\begin{equation}
\boldsymbol{W}\equiv\begin{pmatrix}W\\
-W^{*}
\end{pmatrix},\label{eq:SourcVect}
\end{equation}
with  

\begin{equation}
W(\mathbf{r},t)=\delta V(\theta,t)\Psi_{0}(\mathbf{r},t),\label{eq:SourcFunct}
\end{equation}
 incorporates the effect of the modification of the external potential.
Eq. (\ref{eq:BdGennes}) extends the range of applicability of the
BdG approach to non-stationary setups. In its derivation, no adiabatic
approximation has been made for the time-variation of the background.
Note that, formally, our approach allows dealing with quantum-pressure
effects. However, this line will not be explored in the present work,
as we will introduce for $\Psi_{0}(\mathbf{r},t)$ the result obtained
through the scaling method, which incorporates the Thomas-Fermi approximation.

Now, as the analysis focuses on the azimuthal modes, we use the ansatz 

\begin{equation}
\delta\varphi=\frac{1}{\sqrt{2\pi}}\sum_{n}\left(\alpha_{n}e^{in\theta}+\alpha_{-n}e^{-in\theta}\right)\equiv\sum_{n}\delta\varphi_{n}\qquad\textrm{\emph{n}=1,2,...},\label{eq:DeltaPhiModeAnsatzN}
\end{equation}
 and introduce $\delta\Psi(\mathbf{r},t)=\Psi_{0}(\mathbf{r},t)\delta\varphi(\mathbf{r},t)$
into Eq. (\ref{eq:BdGennes}). The use of constant coefficients $\alpha_{n}$
in the ansatz works better as the quotient between the ring width
and the ring length decreases. Hence, its applicability in the considered
setup is sound. Then, after averaging over the radial and vertical
coordinates with the ground-state density $\rho_{0}(r,z,t)$ found
in the previous section, we obtain for the mode amplitudes

\begin{eqnarray}
i\hbar\begin{pmatrix}\dot{\alpha}_{n}\\
\dot{\alpha}_{-n}^{*}
\end{pmatrix} & = & \begin{pmatrix}\varOmega_{n}(t) & \mathcal{K}(t)\\
-\mathcal{K}(t) & -\varOmega_{n}(t)
\end{pmatrix}\begin{pmatrix}\alpha_{n}\\
\alpha_{-n}^{*}
\end{pmatrix}+\begin{pmatrix}\mathcal{V}_{n}(t)\\
-\mathcal{V}_{-n}^{*}(t)
\end{pmatrix}\qquad\textrm{\emph{n}=1,2,...},\label{eq:alphaEvolution}
\end{eqnarray}
 where 
\begin{eqnarray}
\varOmega_{n}(t) & = & \frac{\hbar^{2}}{2M}n^{2}\left\langle \frac{1}{r^{2}}\right\rangle +g\left\langle \rho_{0}\right\rangle \qquad\textrm{\emph{n}=1,2,...},\label{eq:OMEGAn}\\
\mathcal{K}(t) & = & g\left\langle \rho_{0}\right\rangle ,\label{eq:Kcoupling}\\
\mathcal{V}_{n}(t) & = & \frac{1}{\sqrt{2\pi}}\int_{0}^{2\pi}d\theta\delta V(\theta,t)e^{-in\theta}\qquad\textrm{\emph{n}=1,2,...}.\label{eq:Fourier}
\end{eqnarray}
 The averaged terms are given by 
\[
\left\langle \frac{1}{r^{2}}\right\rangle =\frac{2\pi}{N}\int_{-\infty}^{\infty}dz\int_{0}^{\infty}rdr\frac{1}{r^{2}}\rho_{0}(r,z,t),
\]
 and 
\[
\left\langle \rho_{0}\right\rangle =\frac{2\pi}{N}\int_{-\infty}^{\infty}dz\int_{0}^{\infty}rdr\rho_{0}^{2}(r,z,t),
\]
 where we have taken into account the normalization of the density,
i.e., 
\[
\int_{0}^{2\pi}d\theta\int_{-\infty}^{\infty}dz\int_{0}^{\infty}rdr\rho_{0}(r,z,t)=N.
\]
 Time dependence enters the averages through the evolved ground-state
density. The averaging over the radial and vertical coordinates is
justified given the strong confinement in those directions. Here,
it is worth stressing that our approach corresponds to a time-dependent
variational method. Indeed, Eq. (\ref{eq:BdGennes}) can be derived
from a least-action principle \cite{key-fedichev}. Moreover, Eqs.
(\ref{eq:alphaEvolution}) match the Euler-Lagrange equations obtained
by introducing the ansatz proposed for $\delta\Psi(\mathbf{r},t)=\Psi_{0}(\mathbf{r},t)\delta\varphi(\mathbf{r},t)$
into the corresponding action functional. That framework consistently
incorporates the averaging over the radial and vertical coordinates.

From Eq. (\ref{eq:alphaEvolution}), it is apparent that there is
no coupling between modes with different index $n$. This is a consequence
of the axial symmetry: the expansion does not mix azimuthal modes.
A parallel treatment of the different modes is then feasible. It is
also evident that there are no differential aspects associated to
specific values of $n$. Therefore, without loss of generality, the
analysis can be focused on a particular $n$-mode. Accordingly, in
the following, the ansatz in Eq. (\ref{eq:DeltaPhiModeAnsatzN}) will
be replaced by

\begin{equation}
\delta\varphi_{n}=\frac{1}{\sqrt{2\pi}}\left(\alpha_{n}e^{in\theta}+\alpha_{-n}e^{-in\theta}\right).\label{eq:DeltaPhiModeAnsatzSingleN-1}
\end{equation}

\subsection{Hamiltonian description of the evolution of the mode amplitude}

Significant advances in the analytical characterization of the dynamics
can be achieved using an effective Hamiltonian approach to the evolution
of the mode amplitudes. It is shown that Eq. (\ref{eq:alphaEvolution})
can be derived from the classical Hamiltonian 

\begin{eqnarray}
\mathcal{H}_{n} & = & \varOmega_{n}(t)(\alpha_{n}^{*}\alpha_{n}+\alpha_{-n}^{*}\alpha_{-n})+\mathcal{K}(t)(\alpha_{n}^{*}\alpha_{-n}^{*}+\alpha_{n}\alpha_{-n})+\nonumber \\
 &  & \left[\mathcal{V}_{n}(t)(\alpha_{n}^{*}+\alpha_{-n})+\textrm{c.c.}\right]\label{eq:Hn}
\end{eqnarray}
 through the \emph{Hamilton equations} $i\hbar\dot{\alpha}_{n}=\frac{\partial\mathcal{H}_{n}}{\partial\alpha_{n}^{*}}$
and $i\hbar\dot{\alpha}_{-n}^{*}=-\frac{\partial\mathcal{H}_{n}}{\partial\alpha_{-n}}$.
(In the following, we will take $\hbar=1$) {[}Note that, since $\delta V(\theta,t)$
is real, $\mathcal{V}_{-n}^{*}(t)=\mathcal{V}_{n}(t)${]} In this
framework, we present now a separate treatment of the excitation of
the modes and of their evolution during the expansion.

\subsubsection{The excitation process}

In the experiments, the perturbing potential $\delta V$, applied
to excite the azimuthal modes, was (temporarily) incorporated through
a rectangular ramp. Then, the expansion of the trap was initiated.
We can account for the excitation process with the previously derived
effective Hamiltonian. Since there is no population in the modes when
$\delta V$ is (sharply) introduced, the initial conditions are 

\begin{equation}
\alpha_{n}(0)=0\quad\alpha_{-n}(0)=0.\label{eq:AlphaCInitial}
\end{equation}
 Moreover, as the expansion has not started by that time, there is
no time dependence in $V_{ex}$ and neither in $\Psi_{0}$. Hence,
in Eq. (\ref{eq:Hn}), $\varOmega_{n}$ and $\mathcal{K}$ are constant
coefficients. They will be denoted as $\varOmega_{n}^{(0)}$ and $\mathcal{K}^{(0)}$.
$\mathcal{V}_{n}$ is also constant since, once applied, $\delta V$
does not vary till its (abrupt) cut off. Consequently, the description
of the condensate perturbation simplifies considerably: the (undriven)
Hamiltonian $\mathcal{H}_{n}$ corresponds to two coupled harmonic
modes $\alpha_{n}$ and $\alpha_{-n}$, displaced by constant terms
$\mathcal{V}_{n}$. Defining the variables $\beta_{n}$ and $\beta_{-n}$
through the relations

\begin{equation}
\alpha_{n}=\beta_{n}-\frac{\mathcal{V}_{n}}{\varOmega_{n}^{(0)}+\mathcal{K}^{(0)}}\qquad\alpha_{-n}=\beta_{-n}-\frac{\mathcal{V}_{n}^{*}}{\varOmega_{n}^{(0)}+\mathcal{K}^{(0)}},\label{eq:AlphaBetha}
\end{equation}
 the effective Hamiltonian $\mathcal{H}_{n}$ is rewritten as 

\begin{equation}
\tilde{\mathcal{H}}_{n}=\varOmega_{n}^{(0)}(\beta_{n}^{*}\beta_{n}+\beta_{-n}^{*}\beta_{-n})+\mathcal{K}^{(0)}(\beta_{n}^{*}\beta_{-n}^{*}+\beta_{n}\beta_{-n}),\label{eq:HnAccent}
\end{equation}
 where constant increments in $\tilde{\mathcal{H}}_{n}$ have been
omitted. Then, with the pertinent initial conditions, derived from
Eqs. (\ref{eq:AlphaCInitial}) and (\ref{eq:AlphaBetha}), one can
trivially obtain the evolution of $\beta_{n}$ and $\beta_{-n}$,
and, in turn, that of the original variables $\alpha_{n}$ and $\alpha_{-n}$.
The hold time of the applied rectangular ramp determines the population
of the generated mode: specific values of the mode amplitudes $\alpha_{n}$
and $\alpha_{-n}$ can be achieved by (suddenly) disconnecting the
perturbing potential $\delta V(\theta,t)$ (and, therefore, canceling
the coefficients $\mathcal{V}_{n}$), at appropriate times. Indeed,
by changing the duration interval of the perturbation, a whole range
of mode populations can be reached. In the experimental setup, a sinusoidal
perturbing potential $\delta V(\theta)=\epsilon\sin(n\theta)$ was
applied. Consequently, a mode with a sinusoidal profile, i.e., with
$\alpha_{n}=-\alpha_{-n}$, can be shown to be excited.

\subsubsection{The effect of the ring expansion on the evolution of the mode amplitude}

Once the source potential is disconnected and the expansion of the
ring is made to start, the dynamics of the variables $\alpha_{n}$
and $\alpha_{-n}$ is governed by the Hamiltonian

\begin{equation}
\mathcal{H}_{n}=\varOmega_{n}(\alpha_{n}^{*}\alpha_{n}+\alpha_{-n}^{*}\alpha_{-n})+\mathcal{K}(\alpha_{n}^{*}\alpha_{-n}^{*}+\alpha_{n}\alpha_{-n}),\label{eq:HnUndriven}
\end{equation}
 where $\varOmega_{n}$ and $\mathcal{K}$ are now time-dependent
coefficients given respectively by Eqs. (\ref{eq:OMEGAn}) and (\ref{eq:Kcoupling}).

Using the real variables $J_{n}$, $J_{-n}$, $\eta_{n}$, and $\eta_{-n}$,
defined through the relations

\begin{equation}
J_{n}=\left|\alpha_{n}\right|^{2}\qquad J_{-n}=\left|\alpha_{-n}\right|^{2},\label{eq:RealVariables1}
\end{equation}

\begin{equation}
\eta_{n}=-\arg\left\{ \alpha_{n}\right\} \qquad\eta_{-n}=-\arg\left\{ \alpha_{-n}\right\} ,\label{eq:RealVariables2}
\end{equation}
 the Hamiltonian is rewritten in terms of action-angle variables as 

\begin{equation}
\mathcal{H}_{n}=\varOmega_{n}(J_{n}+J_{-n})+2\mathcal{K}\sqrt{J_{n}J_{-n}}\cos(\eta_{n}+\eta_{-n}).\label{eq:HnActionangle}
\end{equation}

As previously indicated, we work with well-defined initial conditions
which fulfill $\alpha_{n}(0)=-\alpha_{-n}(0)$. (We have redefined
the time origin: $t=0$ corresponds now to the initial time for the
trap expansion). 

Convenient for simplifying the description of the time evolution is
the application of the canonical transformation defined by the generatrix
function 

\begin{equation}
F=\frac{1}{2}\left[J_{n}^{+}(\eta_{n}+\eta_{-n})+J_{n}^{-}(\eta_{n}-\eta_{-n})\right].\label{eq:Generatrix}
\end{equation}
Using the corresponding generalized coordinates 

\begin{equation}
J_{n}^{+}=J_{n}+J_{-n}\qquad J_{n}^{-}=J_{n}-J_{-n}\label{eq:NewActions}
\end{equation}

\begin{equation}
\eta_{n}^{+}=\frac{1}{2}(\eta_{n}+\eta_{-n})\qquad\eta_{n}^{-}=\frac{1}{2}(\eta_{n}-\eta_{-n}),\label{eq:NewAngles}
\end{equation}
 the Hamiltonian is converted into 

\begin{equation}
\mathcal{H}_{n}=\varOmega_{n}J_{n}^{+}+\mathcal{K}\sqrt{J_{n}^{+2}-J_{n}^{-2}}\cos(2\eta_{n}^{+}),\label{eq:HamiltonianPlus}
\end{equation}
 and, from the Hamilton equations, we find 

\begin{equation}
\dot{J_{n}^{-}}=0,\qquad\dot{\eta_{n}^{-}}=-\mathcal{K}\frac{J_{n}^{-}}{\sqrt{J_{n}^{+2}-J_{n}^{-2}}}\cos(2\eta_{n}^{+}),\label{eq:HamilEq1}
\end{equation}

\begin{equation}
\dot{J_{n}^{+}}=2\mathcal{K}\sqrt{J_{n}^{+2}-J_{n}^{-2}}\sin(2\eta_{n}^{+}),\qquad\dot{\eta_{n}^{+}}=\varOmega_{n}+\mathcal{K}\frac{J_{n}^{+}}{\sqrt{J_{n}^{+2}-J_{n}^{-2}}}\cos(2\eta_{n}^{+}).\label{eq:HamilEq2}
\end{equation}
From these equations, it follows that $J_{n}^{-}(t)$ is a constant
of motion, which, for our initial conditions, vanishes. In turn, from
Eqs. (\ref{eq:HamilEq1}), (\ref{eq:NewAngles}), and (\ref{eq:AlphaCInitial}),
it is found that $\eta_{n}^{-}(t)=\eta_{n}^{-}(0)=-\pi/2$.  Making
use of the constant of motion, the evolution of the mode wave function
is expressed as 

\begin{eqnarray}
\delta\varphi_{n}(t) & = & \frac{1}{\sqrt{2\pi}}\sqrt{\frac{J_{n}^{+}(t)}{2}}e^{-i\eta_{n}^{+}(t)}\left(e^{i(n\theta+\pi/2)}+e^{-i(n\theta+\pi/2)}\right)\nonumber \\
 & = & \frac{1}{\sqrt{\pi}}\sqrt{J_{n}^{+}(t)}e^{-i\eta_{n}^{+}(t)}\sin(n\theta).\label{eq:DeltaPhi}
\end{eqnarray}
 Moreover, inserting the above equation into Eq. (\ref{eq:DeltaPhasePsi}),
one finds for the perturbed phase 

\begin{eqnarray}
\delta S(\theta,t) & =- & \frac{1}{\sqrt{\pi}}\sqrt{J_{n}^{+}(t)}\sin[\eta_{n}^{+}(t)]\sin(n\theta)\nonumber \\
 & \equiv & -\chi_{n}(t)\sin(n\theta),\label{eq:DeltaSPhase}
\end{eqnarray}
 where, as convenient for forthcoming discussion, we have singled
out the amplitude $\chi_{n}(t)$. Using the same procedure in Eq.
(\ref{eq:DeltaRhoPsi}), and, after averaging over radial and vertical
coordinates, the perturbed density is found to be given by

\begin{eqnarray}
\delta\rho(\theta,t) & = & \frac{N}{\pi^{3/2}}\sqrt{J_{n}^{+}(t)}\cos[\eta_{n}^{+}(t)]\sin(n\theta)\nonumber \\
 & \equiv & \delta n_{n}(t)\sin(n\theta),\label{eq:DeltaSRho}
\end{eqnarray}
with $\delta n_{n}(t)$ standing for the associated amplitude. In
the following, we will see that the above equations, which give a
complete description of the dynamics of the modes, allow improving
the understanding of the experimental results.

\section{Characterization of the redshift and the Hubble friction}

In Ref. \cite{key-SpielmanRing}, the expansion was observed to induce
the redshift of the mode frequency and the appearance of an effective
damping term, identified as a parallel of the (cosmological) Hubble
\emph{friction}. In the model applied to explain those results, the
evolution of the excited azimuthal modes was described using an effective
space-time metric derived from the background state in the hydrodynamic
formalism. Specifically, it was the evolution of the amplitude $\chi_{n}(t)$
of the perturbed phase that was theoretically characterized: a second-order
differential equation corresponding to a \emph{damped} harmonic oscillator
was obtained for $\chi_{n}(t)$. In the derivation of the effective
metric, the ground state density was assumed to follow the instantaneous
Thomas-Fermi distribution, i.e., an adiabatic approximation for the
substrate was applied. In that framework, the experimental findings
were partially reproduced. The redshift and the Hubble-friction analog
were traced. However, difficulties in precisely reproducing the experimental
value of the effective friction constant were reported \cite{key-SpielmanRing}.
In our study, $\chi_{n}(t)$ is identified from Eq. (\ref{eq:DeltaSPhase})
as 
\begin{equation}
\chi_{n}(t)=\frac{1}{\sqrt{\pi}}\sqrt{J_{n}^{+}(t)}\sin[\eta_{n}^{+}(t)].\label{eq:PhaseAmplitude}
\end{equation}
 Moreover, from Eq. (\ref{eq:DeltaSRho}), it follows that the amplitude
of the perturbed density is 

\begin{equation}
\delta n_{n}(t)=\frac{N}{\pi^{3/2}}\sqrt{J_{n}^{+}(t)}\cos[\eta_{n}^{+}(t)]\label{eq:DensityAmplitude}
\end{equation}

In order to compare with the results presented in \cite{key-SpielmanRing},
we have used the equations of motion, {[}Eqs. (\ref{eq:HamilEq1})
and (\ref{eq:HamilEq2}){]}, to characterize the dynamics of the phase
amplitude $\chi_{n}(t)$. From those equations, we have obtained the
differential equation

\begin{equation}
\ddot{\chi}_{n}-\frac{\dot{\varOmega}_{n}+\dot{\mathcal{K}}}{\varOmega_{n}+\mathcal{K}}\dot{\chi}_{n}+(\varOmega_{n}^{2}-\mathcal{K}^{2})\chi_{n}=0,\label{eq:DifferEquationChi}
\end{equation}
 which can be regarded as corresponding to a harmonic oscillator with
a time-dependent frequency 
\begin{equation}
\omega_{n}(t)=\sqrt{\varOmega_{n}^{2}(t)-\mathcal{K}^{2}(t)},\label{eq:Omega}
\end{equation}
 and an effective \emph{friction} term with time-dependent \emph{damping
}coefficient 
\begin{equation}
\Gamma_{H}(t)=-\frac{\dot{\varOmega}_{n}(t)+\dot{\mathcal{K}}(t)}{\varOmega_{n}(t)+\mathcal{K}(t)}.\label{eq:Gamma}
\end{equation}
 From the Hamiltonian character of the dynamics, it is clear that
there is no dissipation in the system. Terms like \emph{friction }or
\emph{damping} are used in this context simply to emphasize the decrease
in the energy of the perturbation, which, in fact, results from the
driven expansion. 

The analysis of the functional forms of $\Gamma_{H}(t)$ and $\omega_{n}(t)$,
requires evaluating the averages over radial and vertical coordinates
present in Eqs. (\ref{eq:OMEGAn}) and (\ref{eq:Kcoupling}). Using
our analytical results for the evolved ground-state function, we have
obtained 

\begin{eqnarray}
\varOmega_{n}(t) & \simeq & C_{1}r_{c}^{-2}(t)+C_{2}(t)r_{c}^{-\nu}(t),\label{eq:OmegaAveraged}\\
\mathcal{K}(t) & \simeq & C_{2}(t)r_{c}^{-\nu}(t),\label{eq:Kaveraged}
\end{eqnarray}
 where 
\begin{eqnarray}
C_{1} & = & \frac{n^{2}}{2M},\label{eq:C1}\\
C_{2}(t) & = & \frac{1}{3\pi}\sqrt{2gM\omega_{r,eff}(t)\omega_{z,eff}(t)N}\simeq\frac{1}{3\pi}\sqrt{2gM\omega_{r}\omega_{z}N}=C_{2}(0),\label{eq:C2}\\
\nu & = & 1/2\label{eq:NuCoeff}
\end{eqnarray}
 (The validity of the approximation made in Eq. (\ref{eq:C2}) can
be checked from Figs. 3 and 4). In turn, we have found that the dominant
behaviors of the mode frequency and the effective damping \emph{coefficient}
are respectively given by

\begin{eqnarray}
\omega_{n}(t) & \simeq & \sqrt{2C_{1}C_{2}(0)r_{c}^{-(2+\nu)}(t)+\left(C_{1}r_{c}^{-2}(t)\right)^{2}},\label{eq:ModeFrequencyHarmonic}\\
\Gamma_{H}(t) & \simeq & \left(\nu+(2-\nu)\frac{C_{1}}{2C_{2}(0)}r_{c}^{\nu-2}(t)\right)\frac{\dot{r}_{c}(t)}{r_{c}(t)}\equiv\gamma_{H}\frac{\dot{r}_{c}(t)}{r_{c}(t)}.\label{eq:dampingHarmonic}
\end{eqnarray}

Note that our approach has allowed using $r_{c}(t)=R(t)+\lambda_{r}(t)$,
instead of $R(t),$ to incorporate the radial dependence of our results.
This actually conveys a more precise characterization of the studied
effects.

\subsection{The time-dependent frequency}

Eq. (\ref{eq:ModeFrequencyHarmonic}) is the dynamical counterpart,
valid in any regime of time variation, of the dispersion relation
of the static system. In it, the redshift of the mode frequency is
apparent: as the expansion proceeds, (i.e., as $r_{c}$ grows), both
contributions to $\omega_{n}(t)$, namely, the phononic term $2C_{1}C_{2}r_{c}^{-(2+\nu)}(t)$
and the \emph{superluminical} component $\left(C_{1}r_{c}^{-2}(t)\right)^{2}$,
diminish. Given its slower decrease, the phononic term becomes dominant
as the expansion goes on. When the \emph{superluminical} component
can be neglected, the frequency behaves as $\omega_{n}\sim r_{c}^{-(2+\nu)/2}$.
With the obtained value $\nu=1/2$, we find $\omega_{n}\sim r_{c}^{-1.25}$,
close to the the experimental result. As will be shown in the next
section, the inclusion of the anharmonic confinement improves the
agreement. 

Additional differential effects of the \emph{superluminical} component
are apparent in the dependence of $\omega_{n}(t)$ on the mode index.
(Note that $n$ enters Eq. (\ref{eq:ModeFrequencyHarmonic}) through
$C_{1}$ {[}see Eq. (\ref{eq:C1}){]}). The linear behavior, rooted
in the phononic term appears corrected by the $n^{2}$-dependence.
Absent in the expression for the effective frequency obtained in \cite{key-SpielmanRing},
the quadratic dependence appears here because of the more complete
description of the modes. 

The above findings, as those obtained for the radial and vertical
excitations in Sec. III, are relevant beyond the expansion scenario.
For instance, the mere particularization of the above expressions
to the case of a constant central radius is interesting in itself:
the results imply significant advances in the characterization of
the axial modes in a static ring trap. Indeed, that application of
the study can be considered as a parallel of the work presented in
Ref. \cite{key-zaremba} on the properties of the longitudinal modes
in a (static) elongated condensate.

\subsection{The effective friction term}

The appearance of the quotient $\frac{\dot{r}_{c}(t)}{r_{c}(t)}$
in Eq. (\ref{eq:dampingHarmonic}) establishes the connection with
the (cosmological) Hubble \emph{friction}. In the theory presented
in \cite{key-SpielmanRing}, the friction was found to have the form
$\Gamma_{H}(t)=\gamma_{H}\frac{\dot{R}(t)}{R(t)}$. However, the proposed
model was unable to precisely reproduce the experimental findings
with the obtained value of the damping parameter, $\gamma_{H}=1$.
In order to explain the results, the friction was remodeled \emph{ad
hoc} by phenomenologically modifying the damping term. With that variation
of the model, some estimates of the optimal value of $\gamma_{H}$
were made. The best fit to the experimental results was found to correspond
to $\gamma_{H}=0.55$. The structure of the friction term uncovered
by our approach dispenses us from introducing phenomenological elements
in the description: our results fit quite well the experimental findings,
even in the considered case of harmonic trapping. Indeed, retaining
only the dominant term in Eq. (\ref{eq:dampingHarmonic}), we obtain
the approximate expression $\gamma_{H}=\nu=1/2$. Since $C_{1}$ grows
with $n^{2}$, the correction to $\gamma_{H}=1/2$, given by the term
$(2-\nu)\frac{C_{1}}{2C_{2}}r_{c}^{\nu-2}(t)$, can be neglected here
given the small value of the index $n$ for the modes excited in the
experimental realization. 

Useful insight into the actual meaning of the effective friction is
obtained by considering the adiabatic limit in the (driven) dynamics
of the phase amplitude. The existence of an adiabatic invariant was
already apparent in the theoretical framework applied in \cite{key-SpielmanRing}.
In our description, using the reduced Hamiltonian derived by introducing
the constant of motion $J_{n}^{-}(t)=0$ into Eq. (\ref{eq:HamiltonianPlus}),
we find that the (adiabatic-invariant) action $I_{A,n}$ is 

\begin{equation}
I_{A,n}=\frac{E_{n}}{\omega_{n}},\label{eq:AdiabInvarAction}
\end{equation}
 where $E_{n}$, given by

\begin{equation}
E_{n}=(\varOmega_{n}-\mathcal{K})\chi_{n}^{2}+(\varOmega_{n}+\mathcal{K})\delta\zeta_{n}^{2},\label{eq:EffectiveEnergy}
\end{equation}
 is the analog of the energy for the mode-amplitude dynamics. {[}In
the above equation, we have introduced the reduced density amplitude
$\delta\zeta_{n}(t)\equiv\delta n_{n}(t)(N/2\pi)^{-1}${]}. Additionally,
expressing the \emph{energy} as a function of the maximum value $\chi_{n,max}$
of the phase amplitude, the adiabatic invariant is given by 

\begin{eqnarray}
I_{A,n} & = & =\sqrt{\frac{\varOmega_{n}(t)-\mathcal{K}(t)}{\varOmega_{n}(t)+\mathcal{K}(t)}}\chi_{n,max}^{2}(t)=\frac{\chi_{n,max}^{2}(t)}{\sqrt{1+2\frac{C_{2}}{C_{1}}r_{c}^{2-\nu}(t)}},\label{eq:AdiabaInvPhase}
\end{eqnarray}
 which generalizes the expression obtained in \cite{key-SpielmanRing}.
More direct insight is obtained by rewriting $I_{A,n}$ in terms of
the maximum value of the reduced density amplitude $\delta\zeta_{n,max}$

\begin{eqnarray}
I_{A,n} & = & \sqrt{\frac{\varOmega_{n}(t)+\mathcal{K}(t)}{\varOmega_{n}(t)-\mathcal{K}(t)}}\delta\zeta_{n,max}^{2}(t)=\sqrt{1+2\frac{C_{2}}{C_{1}}r_{c}^{2-\nu}(t)}\delta\zeta_{n,max}^{2}(t).\label{eq:AdiabaInvRho}
\end{eqnarray}
 Here, it is apparent that, as the expansion proceeds, since the factor
$\sqrt{1+2\frac{C_{2}}{C_{1}}r_{c}^{2-\nu}(t)}$ grows, the amplitude
of the density perturbation decreases; consequently, the mode tends
to vanish. Although the decrease of the mode \emph{energy}, linked
to that of the effective frequency by Eq. (\ref{eq:AdiabInvarAction}),
can be operatively interpreted in terms of an effective friction,
one must be aware that the system is non-dissipative. The effective
damping is actually induced by the driving of the system incorporated
in $R(t)$. More precisely, it is the expansion that generates the
\emph{energy} decrease. This non-dissipative origin of the effect
becomes evident when the contraction of the ring is considered: in
that variant of the setup, the mode \emph{energy} can be predicted
to grow. 

A comment on the effect of the substrate nonadiabaticity on the mode
dynamics is in order. Here, it is worth recalling that the averages
present in the above analysis have been carried out using the background
density obtained in Sec. III, which is valid irrespective of the regime
of time variation. We have found that, although the similar magnitude
of some of the characteristic times involved in the dynamics is an
argument against the assumption of adiabaticity, some aspects of the
system evolution permit a less restrictive interpretation of the adiabaticity
criteria. Namely, whereas, in the radial dynamics, an approach with
no time-regime limitations is absolutely needed to account for the
displacement of the condensate from the trap center, in the analysis
of the azimuthal excitations, an adiabatic approximation for the substrate
is shown to already reproduce salient features of the modes. Indeed,
the forms obtained for $\omega_{n}(t)$ and $\Gamma_{H}(t)$ with
an adiabatic approach are similar to those given respectively by Eqs.
(\ref{eq:ModeFrequencyHarmonic}) and (\ref{eq:dampingHarmonic}):
the only modification is the replacement of the function $C_{2}(t)$
by the (constant) coefficient $C_{2}=\frac{1}{3\pi}\sqrt{2gM\omega_{r}\omega_{z}N}$.
That approximate agreement is understood taking into account the meaning
of the averaging over the radial and vertical coordinates, applied
in the characterization of the modes. That coarse-graining tends to
smooth out the (oscillatory) background dynamics. Hence, the strong
confinement in the radial and vertical directions, which guarantees
the feasibility of the averaging, implies the attenuation of the substrate
nonadiabaticity when entering the azimuthal mode dynamics. The resulting
blurring of the oscillations justifies to approximate the substrate
evolution as an effective following of the trap. That explains the
satisfactory global picture of the (adiabatic) description given in
\cite{key-SpielmanRing}.

\section{Analysis of the role of the trap anharmonicity in the system dynamics}

In the model used in the previous sections, a harmonic confinement
has been considered. Now, we turn to assess the emergence of differential
effects linked to the trap anharmonicity.

\subsection{The evolution of the ground state in an anharmonic trap}

Since the nonlinear confinement cannot be tackled with the scaling
methods, we must deal with numerically solving the GP equation with
the complete trapping potential present in Eq. (\ref{eq:ExternalPotential}).
Again, we have used the ansatz given by Eq. (\ref{eq:NonPolyAnsatz})
and have solved Eqs. (\ref{eq:NonPolyn}) and (\ref{eq:NonPolyWidth}).
We stress that the nonlinear confining potential in the radial direction
is straightforwardly incorporated in the applied method. The numerical
results depicted in Fig. 5 show that the short-term oscillatory radial
displacement of the condensate presents characteristics similar to
those of the harmonic case. In contrast, at larger times, differences
are observed, in particular, as the damping of the oscillations sets
in. This is actually a dephasing of the wavepackect components due
to  the anharmonic confinement. Additionally, a more irregular evolution
of the condensate shape is observed, as shown in Fig. 6. This is particularly
evident after the expansion. Because of the deformation, the evolution
cannot be regarded now as corresponding to a radial dipole mode. The
complex character of the post-inflationary dynamics detected in the
numerical simulations of \cite{key-SpielmanRing} is already apparent
in our effective mono-dimensional description: it can be interpreted
as resulting from the reflection  of the displaced state, generated
by the radial translation during the expansion, from the \emph{walls}
of the (sharp) anharmonic potential. The energy of the state, determined
by the displacement at the end of the expansion, grows with the velocity
of the radius ramp. Previous studies \cite{key-fromhold,key-schleich,key-leadbeater}
on the generation of solitons in related contexts provide useful clues
to this picture of the \emph{preheating} stage. From them, one can
conjecture the key importance of reaching significant values of the
ramp velocities, and, therefore, of the state energy, to the emergence
of solitons. The precise characterization of the production of solitons
and vortices is left for future work. 

\begin{figure}[H]
\centerline{\includegraphics{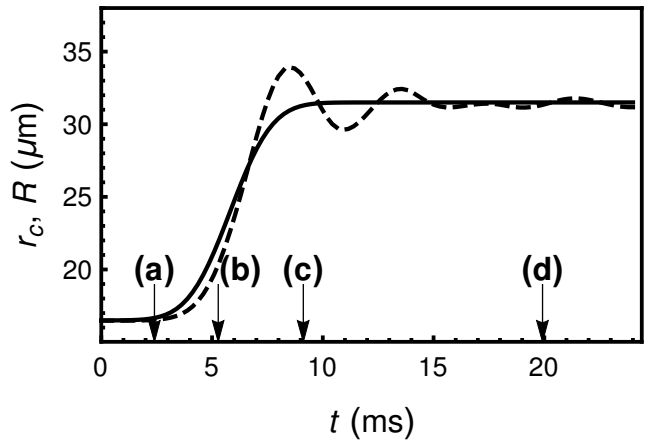}}\caption{The trap radius $R(t)$ (continuous line) and the radial position
of the average value of the radial coordinate of the condensate $r_{c}(t)$
(dashed line) as functions of time. The form of the radius ramp is
the same as that used in \cite{key-SpielmanRing}.  The arrows correspond
to times shown in Fig. 6. The system parameters are the same as those
used in Fig. 1; additionally, $\Lambda/h=0.8\,\textrm{Hz\ensuremath{\mu m^{-4}}}$.}
\end{figure}

\begin{figure}[H]
\centerline{\includegraphics{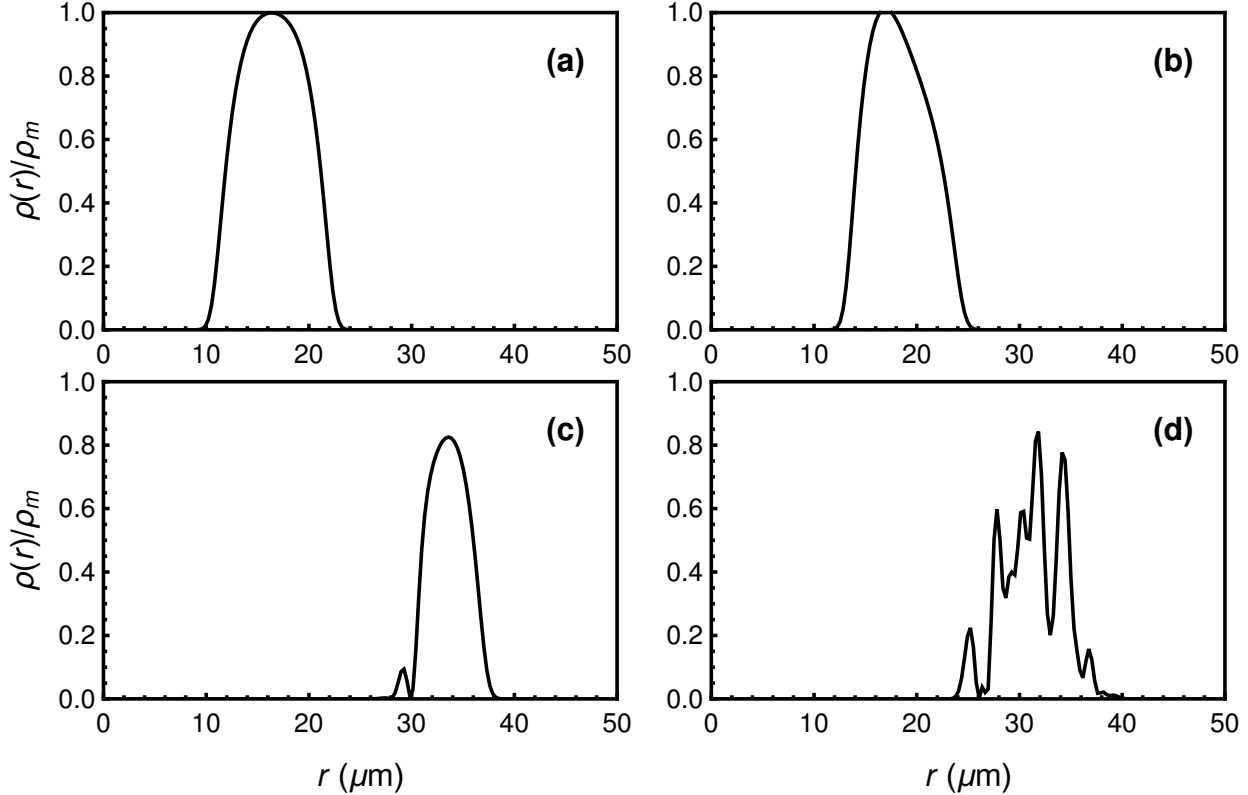}}\caption{The reduced radial density $\rho(r)/\rho_{m}$ at four different times.
($\rho_{m}$ denotes the maximum of $\rho(r)$ before the expansion).
The labels $a$, $b$, $c$, and $d$ correspond to the times identified
in Fig. 1. The system parameters are the same as those used in Fig.
5. The results were obtained by numerically solving the non-polynomial
GP equation. The system parameters are the same as those used in Fig.
5.}
\end{figure}

\subsection{Trap-anharmonicity corrections to the azimuthal-mode dynamics}

In order to evaluate the effect of the trap nonlinearity on the dynamics
of the azimuthal modes, we have repeated the procedure used in Sec.
V to derive the differential equation for $\chi_{n}(t)$ considering
now the quartic potential $\Lambda[r-R(t)]^{4}$ and applying an adiabatic
approach for the evolved ground state.  Since the inclusion of the
(sharp) anharmonic term increases the radial confinement, the quality
of the averaging increases, and, consequently, a higher accuracy of
the adiabatic approximation can be expected. 

The obtained expressions for the mode frequency $\omega_{n}(t)$ and
for the effective friction term $\Gamma_{H}(t)$ have the same form
as their \emph{harmonic} counterparts, given in Eqs. (\ref{eq:ModeFrequencyHarmonic})
and (\ref{eq:dampingHarmonic}). Only, the characteristic parameters
differ: the previously given \emph{harmonic} parameters $C_{2}$ and
$\nu$ are replaced now by the \emph{anharmonic} ones 
\begin{eqnarray}
C_{2}^{(an)} & = & \frac{1}{11}\left(\frac{2}{\pi^{2}}\right)^{3/7}\lambda^{1/7}\left(\frac{21g\sqrt{M}N\omega_{z}\Gamma(3/4)}{\Gamma(1/4)}\right)^{4/7},\label{eq:AnharmC2}\\
\nu^{(an)} & = & 4/7.\label{eq:AnharmNu}
\end{eqnarray}

With the obtained value of $\nu^{(an)}$, we find for the effective
frequency $\omega_{n}\sim R^{-(1+2/7)}$, in complete agreement with
the experimental results. Moreover, the friction coefficient is found
to be $\gamma_{H}=\nu^{(an)}=4/7$, close to the estimate of the value
that best reproduces the experimental results, $\gamma_{H}=0.55$,
and significantly different from $\gamma_{H}=1$, derived with the
model presented in Ref. \cite{key-SpielmanRing}. Once more, it is
apparent that the introduction of phenomenological damping to fit
the experimental findings becomes unnecessary in our approach.

\section{Concluding remarks}

The presented analytical characterization of the different mechanisms
relevant to the inflationary dynamics of the condensate constitutes
a useful tool in the analysis of the experiments of Ref. \cite{key-SpielmanRing}.
Particularly interesting findings of our study are the oscillatory
character of the radial displacement of the condensate and the coupling
of the radial and vertical widths in the modes excited by the expansion.
Our picture has allowed evaluating the differential role of the radius
ramp in those effects. Whereas the shape scaling functions are directly
affected by $R$, it is the acceleration $\ddot{R}$ that enters as
a driving term the equation for the condensate displacement. Also
noticeable is how the mode coupling, which is specific to the considered
ring geometry and becomes especially important in a contracting setup,
can be modified by changing the relative magnitude of the radial and
vertical frequencies. 

Despite the complex character of the radial dynamics (i.e., of the
inflaton-field analog), it has been possible to trace the origin of
the \emph{preheating} process to two elements: the radial displacement
of the condensate at the end of the radius ramp and the nonlinear
component of the trapping potential. Our analysis can assist the design
of strategies to control the process. For instance, given the indispensable
role of the anharmonicity, the use of a stronger harmonic confinement,
planed for future experiments \cite{key-SpielmanRing}, can be predicted
to suppress the \emph{preheating}. Additionally, the requirement of
working with fast  ramps to observe the complex post-inflationary
dynamics can be linked to the need of having a significant amplitude
in the radial displacement: the generation of solitons via the reflection
of the displaced states in  the \emph{walls} of the anharmonic confinement
requires of sufficiently high energy. It is also concluded that, in
an adiabatic regime, since there is no radial displacement with respect
to the trap, the complex post-inflationary processes are not activated. 

The generalized BdG approach used to analyze the azimuthal modes has
allowed improving the characterization of the red-shifting. Indeed,
the expression found for the dynamical dispersion relation has uncovered
the contribution of the superluminical component, which cannot be
neglected when high-index modes are involved. Also, an extended picture
of the Hubble-friction mechanism has been given. As a result, a precise
fit of the effective damping constant has been achieved without introducing
phenomenological corrections in the basic model. 

The applicability of the scaling approach to generic variations of
the trap radius, (including contractions), can provide a variety of
analytical results on red-shifting (or blue-shifting) and Hubble \emph{friction}
(or acceleration), which can enrich the versatility of the system
as a scenario for \emph{cold-atom cosmology}. Additionally, the particularization
of the study to a constant radius has the general interest of describing
the normal modes in a static ring trap. Methods for the selective
excitation of  modes and for their experimental detection  can be
designed from the given picture. 

In future work, we plan to deal with the generation of solitons and
vortices in the post inflationary dynamics. To this end, the detailed
characterization of the role of the condensate displacement at the
end of the expansion will be tackled. We also intend to evaluate the
relevance  of dynamical instability  to the considered system. Although
their presence in the analyzed inflationary dynamics has been discarded,
the unstable modes can have a significant role in other regimes. In
this line, we plan to explore the possibility of tracing a parallel
in the ring-trap scenario of the mechanism responsible for the dynamical
Casimir effect studied in  Ref. \cite{key-parentaniInflation}. There,
appropriate changes in the transversal frequencies of elongated BECs
were used to modify the effective interaction strength in the longitudinal
direction. The excitation of specific  modes from the vacuum state
via parametric resonance were predicted. In our system, the azimuthal
interaction strength can also be modified through changes in the radial
and vertical dynamics. Hence, it is pertinent to analyze the potential
use of resonances to excite modes from the vacuum and to establish
the connection with other processes  where unstable modes are also
relevant.

\appendix*

\section{EFFECTIVE DECOUPLING OF THE CONDENSATE RADIAL AND VERTICAL WIDTHS}

For $\omega_{z}\gg\omega_{r}$, approximate solutions to the system
of Eqs. (\ref{eq:DeltaSigmaR}) and (\ref{eq:DeltaSigmaZ}) can be
obtained through an adiabatic approximation. (Note that the term adiabatic
in this context has a meaning different from that corresponding to
the dynamics of the background with respect to the trap variation).
First, as $\delta\sigma_{r}$ evolves much more slowly than $\delta\sigma_{z}$,
Eq. (\ref{eq:DeltaSigmaZ}) is solved for a \emph{frozen} value of
$\delta\sigma_{r}$, which will be denoted as $\tilde{\delta\sigma_{r}}$:

\begin{equation}
\ddot{\delta\sigma_{z}}=-\omega_{z}^{2}\left(\tilde{\delta\sigma_{r}}+3\delta\sigma_{z}\right)\label{eq:DeltaSigZ}
\end{equation}
 The \emph{equilibrium} value is then given by 
\begin{equation}
\delta\sigma_{z,eq}=-\frac{\tilde{\delta\sigma_{r}}}{3}\label{eq:DeltaSigZEq}
\end{equation}
 Now, writing 
\begin{equation}
\delta\sigma_{z}=\delta\sigma_{z,eq}+\tilde{\delta\sigma_{z}},\label{eq:DeltaSigZAnsatz}
\end{equation}
 we obtain from Eq. (\ref{eq:DeltaSigZ})

\begin{equation}
\ddot{\tilde{\delta\sigma_{z}}}=-3\omega_{z}^{2}\tilde{\delta\sigma_{z}},\label{eq:DeltaSigZAccent}
\end{equation}
 which is trivially solved to give
\begin{equation}
\tilde{\delta\sigma_{z}}=C\sin(\sqrt{3}\omega_{z}t+\vartheta),\label{eq:DeltaSigZSol}
\end{equation}
 where the constants $C$ and $\vartheta$ are  determined by the
initial conditions. Now, introducing $\delta\sigma_{z}$, as given
by Eqs. (\ref{eq:DeltaSigZAnsatz}) and (\ref{eq:DeltaSigZSol}),
into Eq. (\ref{eq:DeltaSigmaR}), and averaging the higher frequency
signal, one finds

\begin{equation}
\ddot{\delta\sigma_{r}}=-\frac{8}{3}\omega_{r}^{2}\delta\sigma_{r}.\label{eq:DeltaSigR}
\end{equation}
 Hence, from this equation and from Eq. (\ref{eq:DeltaSigZAccent}),
the effective mode frequencies are obtained as $\tilde{\omega}_{r}=\sqrt{8/3}\omega_{r}$
and $\tilde{\omega}_{z}=\sqrt{3}\omega_{z}$.

\section*{Acknowledgments}

One of us (JMGL) acknowledges the support of the Spanish Ministerio
de Economía y Competitividad and the European Regional Development
Fund (Grant No. FIS 2016-79596-P).

\end{document}